\documentclass[10pt]{article}
\usepackage{feynmf}

\begin{document}

\title{The Coupling of $\eta$ Mesons to Quarks
and Baryons from $D_s^*\rightarrow D_s\pi^0$ Decay}

\author{T.A. L\"ahde$\,$\footnote{talahde@pcu.helsinki.fi}$\:$
and$\:\:$D.O. Riska$\,$\footnote{riska@pcu.helsinki.fi} 
\\ \vspace{0.3cm} \\
{\normalsize \it Helsinki Institute of Physics
and Department of Physics}\\
{\normalsize\it PL 64, University of Helsinki, 00014 Finland} }

\date{}
\maketitle
\thispagestyle{empty}

\begin{abstract}
The known ratio of the branching fractions for $D_s^*\rightarrow D_s\pi^0$ 
and $D_s^*\rightarrow D_s\gamma$ may be used to extract the coupling of 
$\eta$ mesons to strange quarks once the value of the $\pi^0\!-\!\eta$ 
mixing angle is known. This requires that realistic models for the spectra 
as well as the magnetic dipole~(M1) decays of the heavy-light~($Q\bar q$) 
mesons are available. The coupling of $\eta$ mesons to light quarks may 
then be estimated using $SU(3)$ flavor symmetry. Applied to the quark 
model for the baryons, an $\eta NN$ pseudovector coupling constant
of $f_{\eta NN}\:=\: 0.35\:^{+0.15}_{-0.25}$ is obtained. If the charm 
quark couples significantly to the $\eta$ meson, as is suggested by the 
decay mode $\psi'\rightarrow J/\psi\,\eta$, then somewhat larger values of 
$f_{\eta NN}$ can be obtained. These values are sufficiently small to be 
consistent with phenomenological analysis of photoproduction of the $\eta$ 
on the nucleon and the reaction $pp\rightarrow pp\eta$. 
\end{abstract}

\newpage

\section{Introduction}

The flavor symmetry breaking $D_s^*\rightarrow D_s\pi^0$ decay mode 
is mainly due to a small isoscalar $\eta$ meson component in the physical
$\pi^0$ meson as only the $\eta$ meson can couple to the strange quark in
the strange-charm mesons~\cite{Wise,Stewart}. As the magnitude of the 
$\pi^0\!-\!\eta$ mixing angle has been at the focus of a considerable 
amount of theoretical and phenomenological analysis, its value is now 
rather well known~\cite{Magiera,Henley}. The expression given by chiral 
perturbation theory to lowest order~\cite{Gasser} is, for small $\theta_m$, 
\begin{equation}
\theta_m\:=\:{\sqrt{3}\over 4}{m_d-m_u\over m_s-m_0}\:\simeq\: 0.010
\label{e1}
\end{equation}
where $m_0$ is the average of the $u$ and $d$ current quark masses. To 
this should be added the contribution from $\pi^0\!-\!\eta'$ mixing, which 
is smaller because of the larger mass of the $\eta'$, and 
which may increase the value of $\theta_m$ by at most about 
30\%~\cite{Magiera,Coon}. The resulting effective $\pi^0\!-\!\eta$ mixing 
angle is therefore about $\sim 0.012$.

Combination of the notion of the spontaneously broken approximate chiral 
symmetry of QCD with $SU(3)$ flavor symmetry implies that the coupling of 
the baryons to the octet of light pseudoscalar mesons ($\pi,K,\eta$) takes 
the form
\begin{equation}
{\cal L}_{N\!N\!M}\:\:=\:\:i\frac{g_A}{2f_m}\:\bar\psi_N\: 
\gamma_5\gamma_\mu\partial_\mu\:m_a\lambda_a\:\psi_N,
\label{lNN}
\end{equation}
where $\psi_N$ denotes the baryon octet field, $m_a$ the meson octet 
field, $\lambda_a$ the $SU(3)$ Gell-Mann matrices, $f_m$ the decay 
constants of the mesons, and $g_A$ the axial coupling constant of the 
baryons. The current experimental value of $g_A$ as determined from low 
energy nuclear beta decays is $g_A=1.2573$~\cite{Wbook}, while the 
phenomenological value is $g_A=1.267\pm 0.004$~\cite{Feldmann}. If these 
values are employed together with the Goldberger-Treiman relations for 
nucleons, then "effective" pseudovector coupling constants may be 
defined as
\begin{equation}
f_{\pi NN} = {m_\pi\over 2 f_\pi}\,g_A \simeq 0.95,\quad\quad
f_{\eta NN}= \frac{1}{\sqrt{3}}\frac{m_\eta}{2f_\eta}\,g_A \simeq 1.79. 
\label{e3}
\end{equation}
The above value for the $\pi NN$ coupling constant agrees completely with 
that used in the most recent realistic phenomenological nucleon-nucleon 
interaction models, the parameters of which have been determined by fits 
to nucleon-nucleon scattering data~\cite{Nijm,V18}. Such fits do not, 
however, constrain the value of $f_{\eta NN}$ very well, and therefore its 
value remains uncertain. Different realistic phenomenological potential 
models give values between $\sim 0.5$~and~$\sim 3.6$ for this 
coupling~\cite{DOR}, while analysis of data on photoproduction of the 
$\eta$ on the nucleon indicate that the value for $f_{\eta NN}$ should not 
exceed 0.64~\cite{Tiator}, a value which also seems to be consistent with 
data on $\eta$ production near threshold in proton-proton 
collisions~\cite{Pena}. 

The point of the present paper is to show that it is possible, given the 
value of the $\pi^0\!-\!\eta$ mixing angle, to estimate the strength of 
the coupling of the $\eta$ meson to strange quarks from the empirically 
known ratio of the branching fractions for $D_s^*\rightarrow D_s\pi^0$ and
$D_s^*\rightarrow D_s\gamma$. This requires that realistic models for the 
respective hadronic matrix elements and $Q\bar q$ wavefunctions are at 
hand. From the 
$\eta ss$ coupling obtained, 
the corresponding $\eta qq$ coupling may be estimated using 
flavor $SU(3)$ symmetry. One may then, by standard quark model relations, 
obtain useful and constraining information on the $\eta-$baryon, and in 
particular, the $\eta-$nucleon coupling $f_{\eta NN}$. The main source of 
error in this analysis is the rather large uncertainty in the empirical 
determination of the $\pi^0$ and $\gamma$ branching fractions for the 
$D_s^*$ meson. This error is likely to be much larger than that introduced 
by $SU(3)$ breaking effects into the current analysis.

If one considers the coupling of the octet of light pseudoscalar mesons to 
the light ($u,d,s$) quarks, one obtains analogously to eq.~(\ref{lNN}) the 
couplings
\begin{equation}
{\cal L}_{qqM}\:\:=\:\:i\frac{g_A^q}{2f_m}\:\bar\psi_q\: 
\gamma_5\gamma_\mu\partial_\mu\:m_a\lambda_a\:\psi_q,
\label{lqq}
\end{equation}
where $g_A^q$ denotes the axial coupling constant for constituent quarks.
For the pions and the $\eta$ meson, the empirical values of the decay 
constants are \mbox{$f_\pi$ = 93 MeV} and \mbox{$f_\eta$ = 112 MeV,} 
respectively, so at least in this case the $SU(3)$ flavor symmetry is 
broken only at the 10\% level. Combination of the chiral 
coupling~(\ref{lqq}) with the representation
\begin{equation}
m_a\lambda_a = \sqrt{2}
       \left[ \begin{array}{ccc}
        \frac{\pi^0}{\sqrt{2}} + \frac{\eta^0}{\sqrt{6}} & \pi^+ & K^+ \\
        \pi^- & -\frac{\pi^0}{\sqrt{2}} + \frac{\eta^0}{\sqrt{6}} & K^0 \\
        \bar K^- & \bar K^0 & -\sqrt{\frac{2}{3}}\eta^0
       \end{array} \right], \quad
\psi_q = \left( \begin{array}{c}
        u \\ d \\ s
       \end{array} \right)
\label{SU3matr}
\end{equation}
leads to the following definitions for the quark-level pseudovector 
coupling constants $f_{mqq}$, which are analogous to those given by 
eq.~(\ref{e3}):
\begin{equation}
f_{\pi qq} = \frac{m_\pi}{2f_\pi}g_A^q, \quad 
f_{\eta qq} = \frac{m_\eta}{2\sqrt{3}f_\eta}g_A^q, \quad
f_{\eta ss} = -\frac{m_\eta}{\sqrt{3}f_\eta}g_A^q.
\label{Treimanq}
\end{equation} 
The above relations then suggest that the magnitude of the coupling of 
$\eta$ mesons to $u,d$ quarks should be one-half that of the $\eta$ 
coupling to strange quarks, independently of the $\eta$ meson mass. The 
numerical value of $g_A^q$ is not very well known, although typical values 
fall in the range 0.87 - 1~\cite{Wein1,Wein2}. Calculations of the pionic 
decay widths of the $D^*$ mesons show that value range to be 
realistic~\cite{pidec,Goity}. In the static quark model the meson-quark 
coupling constants eq.~(\ref{Treimanq}) are related to the meson-nucleon 
coupling constants given by 
eq.~(\ref{e3}) as (see e.g.~ref.~\cite{Goity})
\begin{equation}
f_{\pi NN} = \frac{5}{3}\,f_{\pi qq}, \quad
f_{\eta NN} = f_{\eta qq}.
\label{qmod}
\end{equation}
Thus for $g_A\sim 1.26$ one would predict a value of $g_A^q\simeq 0.75$, 
which may be considered as a lower bound for the constituent quark axial 
coupling~\cite{Goity}. If the relation~(\ref{Treimanq}) is 
employed with values of $g_A^q$ in the range 0.87 - 
1.0~\cite{Wein1,Wein2}, then the corresponding values of $f_{\eta ss}$ 
would fall between -2.5 and -2.8. Note that the value of $g_A^q$ is 
conventionally extracted from the axial (spatial) current coupling, as the 
value of the effective axial charge coupling of confined constituent quarks 
depends on the form of the confining interaction~\cite{pidec}.

The empirical observation of the charmonium decay modes $\psi'\rightarrow 
J/\psi\,\pi^0$ and $\psi'\rightarrow J/\psi\,\eta$ indicates that charm 
quarks also couple to the $\pi^0$, and may potentially influence the 
$D_s^*\rightarrow D_s\pi^0$ decay. The fairly large empirical width for 
$\psi'\rightarrow J/\psi\,\eta$ of $\sim$ 7.5 keV which corresponds to a 
branching fraction of $\sim 3\%$ suggests that besides $q\bar q$ and 
$s\bar s$ the $\eta$ meson may have a $c\bar c$ admixture. While this 
would allow direct coupling of the $\eta$ meson to charm quarks, the charm 
components of the $\eta$ and $\eta'$ mesons are phenomenologically, as well 
as theoretically, constrained to be rather small~\cite{Feldmann, Franz, 
Franz2}. The known decay width for $\psi'\rightarrow 
J/\psi\,\eta$ ~\cite{PDG,Amet} and the Hamiltonian model for the $c\bar c$ 
spectrum of ref.~\cite{Resc} may nevertheless be used to explore the 
consequences of a nonzero $\eta cc$ coupling for the decay 
$D_s^*\rightarrow D_s\pi^0$ and thereby the resulting value for $f_{\eta 
NN}$. For this purpose the effective $\eta cc$ coupling is here taken to 
have the form~(\ref{lqq}).

This paper is divided into 5 sections. In section 2, the method for 
extracting the $\eta ss$ coupling from the ratio of the empirically
known branching fractions for the decays $D_s^*\rightarrow D_s\pi^0$ and
$D_s^*\rightarrow D_s\gamma$ is described, along with the models for the 
$\pi^0$ and $\gamma$ decays of the $D_s^*$. It is shown that a 
relativistic treatment of both the $\gamma$ and $\pi^0$ decays of the 
$D_s^*$ is called for. In section 3 it is shown that the two-quark 
exchange current contributions to the $\gamma$ decay amplitude, which
are associated with the scalar confining and vector one-gluon 
exchange~(OGE) components of the $Q\bar q$ interaction, provide a way of 
obtaining a realistic value for the matrix element for $\gamma$ decay, 
which is not possible in the impulse approximation. These two-quark 
contributions appear here because of the explicit elimination of negative 
energy components in the Blankenbecler-Sugar~(BSLT) reduction of the 
Bethe-Salpeter equation. Section~\ref{res-sec} contains the numerical 
results for the $\pi^0$ and $\gamma$ matrix elements, and an estimated 
upper bound for the $\eta cc$ coupling. Section 5 contains a concluding 
discussion of the results obtained for the coupling of $\eta$ mesons to 
quarks and nucleons.

\vspace{-0.25cm}

\section{The $D_s^*\rightarrow D_s\pi^0$ and $D_s^*\rightarrow D_s\gamma$ 
Decays}
\label{2sec}

\subsection{Extraction of the $\eta ss$ coupling}

Application of the relations~(\ref{Treimanq}) together with 
eq.~(\ref{lqq}) yields a coupling of $\eta$ mesons to strange quarks in 
terms of $f_{\eta ss}$. The coupling of the $\pi^0$ meson to the strange 
quark may then be expressed in terms of the "effective" mixing angle 
$\theta_m$ as \vspace{-0.35cm}
\begin{equation}
{\cal L}_{ss\pi^0}\:\:=\:\:
i\frac{f_{\eta ss}}{m_\eta}\,\theta_m\:
\bar \psi_s\,\gamma_5\gamma_\mu\partial_\mu \phi_{\pi^0}\,\psi_s,
\label{e4}
\end{equation}
where $\sin(\theta_m)$ has been replaced by $\theta_m$, which is a very 
good approximation for values of $\theta_m \sim 0.015$ rad. In 
eq.~(\ref{e4}), $\theta_m$ now corresponds to the sum of the
$\pi^0\!-\!\eta$ and $\pi^0\!-\!\eta'$ contributions as discussed in the 
previous section.

The differential decay width of an excited heavy-light ($Q\bar q$) meson 
may be expressed in the form
\begin{equation}
\frac{d\Gamma}{d\Omega} = \frac{q}{8\pi^2}\frac{M_f}{M_i} |T_{fi}|^2,
\label{dec}
\end{equation}
where $q$ denotes the momentum of the emitted particle, $T_{fi}$ is the 
hadronic decay amplitude for the process in question, and $M_f/M_i$ is a 
normalization factor for the quarkonium states analogous to that employed 
in ref.~\cite{Goity}. The width for the process $D_s^*\rightarrow 
D_s\pi^0$ may then be obtained directly from the expression for 
$D^{\pm*}\rightarrow D^\pm\pi^0$ given in ref.~\cite{pidec} 
by the replacement $g_A^q / 2f_\pi \rightarrow f_{\eta ss} \theta_m / 
m_\eta$, giving
\begin{equation}
\Gamma(D_s^*\rightarrow D_s\pi^0)={1\over 6\pi}
{M_{D_s}\over M_{D_s^*}}{f_{\eta ss}^2\over m_\eta^2}
\,\theta^2_m q_\pi^3\,|{\cal M}_\pi|^2,
\label{e6}
\end{equation} 
if the $\pi^0$ emission takes place at the strange quark.
Here ${\cal M}_\pi$ is a radial matrix element for pion emission. 
Similarly, using eq.~(\ref{dec}) the decay width for the 
radiative M1 transition $D_s^*\rightarrow D_s\gamma$ may be written in the 
form~\cite{M1}
\begin{equation}
\Gamma(D_s^*\rightarrow D_s\gamma) =
{16\over 3}{M_{D_s}\over M_{D_s^*}}\,\alpha
\,q_\gamma^3\, |{\cal M}_\gamma|^2,
\label{e8}
\end{equation}
where $\alpha$ is the fine structure constant. In analogy with 
eq.~(\ref{e6}), ${\cal M}_\gamma$ is a radial matrix element for M1 decay, 
which includes the inverse quark mass factor (see below). By means of 
eqs.~(\ref{e6}) and~(\ref{e8}), the ratio of the $\pi^0$ and 
$\gamma$ decay widths of the $D_s^*$ meson is then obtained as
\begin{equation}
\frac{\Gamma_\pi}{\Gamma_\gamma} = \frac{8}{9\pi} 
\frac{f_{\eta ss}^2\theta^2_m}{m_\eta^2 \alpha} 
\left(\frac{q_\pi}{q_\gamma}\right)^3
\left(\frac{|{\cal M}_\pi|}{|{\cal M}_\gamma|}\right)^2.
\label{e11}
\end{equation}
Note that in the above equation, the dimension of $|{\cal M}_\gamma|$ 
is [MeV]$^{-1}$. Through use of the empirical ratio of pion and photon 
momenta~\cite{PDG} known to be approximately~139/48 and the $\eta$ meson 
mass of 547 MeV one may solve for the coupling constant $f_{\eta ss}$ to 
get
\begin{equation}
f_{\eta ss}^2 = {\theta^{-2}_m}\:\frac{\Gamma_\pi}{\Gamma_\gamma} 
\:\left(\frac{|{\cal M}_\gamma|}{|{\cal M}_\pi|}\right)^2 \cdot
\,4.814\:\mathrm{fm^{-2}}.
\label{rat}
\end{equation}
As the ratio of the $\pi^0$ and $\gamma$ decay rates is experimentally 
known, albeit with quite large errors, to be $0.062\pm 0.028$~\cite{PDG}, 
it is, given the rather well known value of $\theta_m$, possible to obtain 
an estimate for the coupling constant $f_{\eta ss}$. This requires a model 
for the matrix elements in eq.~(\ref{rat}), as discussed below. 


Eq.~(\ref{rat}) was obtained by assuming that the $\pi^0$ decay takes 
place only at the strange quark. However, the observed 
$\psi'\rightarrow J/\psi\,\eta$ decay shows that
the $\eta$ meson also couples to charm quarks, which implies that  
$\pi^0$ emission by the charm quark may also take place in the $D_s^*$ 
meson in addition to $\pi^0$ emission by the strange quark through 
$\pi^0\!-\!\eta$ mixing. If the pion emission at the charm quark is taken 
to proceed through a coupling of the type~(\ref{e4}), then the effect of 
$\pi^0$ emission by the charm quark may be taken into account by 
replacement of eq.~(\ref{rat}) with the expression
\begin{equation}
R = {\theta^{-2}_m}\:
\frac{\Gamma_\pi}{\Gamma_\gamma}\:|{\cal M}_\gamma|^2 \cdot
\,4.814\:\mathrm{fm^{-2}},
\label{Req}
\end{equation}
where $R$ is defined as
\begin{equation}
R = \left|f_{\eta ss}\,{\cal M}^s_\pi + f_{\eta cc}\,{\cal 
M}^c_\pi\right|^2.
\label{req}
\end{equation}
Here ${\cal M}^c_\pi$ denotes the radial matrix element for pion emission 
by the charm quark. Because of the small momentum of the emitted $\pi^0$ 
and the heaviness of the charm quark, this matrix element will be very 
close to~1. The phenomenological implications of a nonzero contribution to 
$\pi^0$ decay by the charm quark will be explored in section~\ref{4sec}.

The coupling constant $f_{\eta cc}$ can only be estimated from the 
empirically observed decay $\psi'\rightarrow J/\psi\,\eta$ which, although 
somewhat suppressed by the orthogonality of the $\psi'$ and $J/\psi$ 
wavefunctions, is known~\cite{PDG} to have a rather large width of $\sim 
7.5$ keV corresponding to a branching fraction of about 3\%. However, 
reliable estimation of the coupling constant $f_{\eta cc}$ from the 
decay $\psi'\rightarrow J/\psi\,\eta$ is difficult for several reasons. 
The first is that the matrix element for $\psi'\rightarrow J/\psi\,\eta$ 
is very small in the nonrelativistic approximation because of the 
orthogonality of the wavefunctions. In a relativistic calculation this 
matrix element is typically increased by an order of magnitude. 
Furthermore, the axial charge component from a coupling of the 
type~(\ref{e4}), along with the associated two-quark effects, may also 
contribute significantly to such a decay~\cite{pidec}. Thus if the axial 
current component of eq.~(\ref{e4}) is used with the replacement $f_{\eta 
ss}\theta_m \rightarrow f_{\eta cc}$, then the value for $f_{\eta cc}$ 
obtained by fitting the empirical width for $\eta$ decay of the 
$\psi'$ should be expected to represent an upper bound only. The 
expression for the width so obtained is
\begin{equation}
\Gamma(\psi'\rightarrow J/\psi\,\eta) = \frac{4}{3\pi}
\frac{M_{J/\psi}}{M_{\psi'}} \frac{f_{\eta cc}^2}{m_\eta^2}\:q_\eta^3\:
|{\cal M}_\eta|^2.
\label{etawidth}
\end{equation}
Here ${\cal M}_\eta$ is the matrix element for $\eta$ decay, which is 
given in the next section together with the matrix element ${\cal M}_\pi$.
Note that in eq.~(\ref{etawidth}), the difference of a factor 8 as 
compared with eq.~(\ref{e6}) arises from the product of a factor 2 from 
the spin sum for a triplet-triplet transition and a factor 4 from the sum 
of the quark and antiquark contributions to $\eta$ decay. In 
section~\ref{res-sec} it is shown that evaluation of eq.~(\ref{etawidth}) 
leads to an upper bound for $f_{\eta cc}$ of about $\sim 0.8$.

\subsection{Matrix elements for pseudoscalar emission}

When the strange and charm constituent quarks are treated 
non-relativistically, the radial matrix element for
$\pi^0$ decay, ${\cal M}_\pi$, is
\begin{equation}
{\cal M}_\pi^{\mathrm{NR}} = \int_0^\infty 
dr\,u^2(r)\:j_0\left(\frac{q_\pi r}{2}\right) 
\simeq 1 - \frac{\left<r^2\right> q_\pi^2}{24} + {\cal O}(q_\pi^4),
\label{nrpimat}
\end{equation}
where $u(r)$ is the reduced radial wavefunction of $c\bar s$ system. 
Because of the smallness of $q_\pi$ (48 MeV) and 
$\left<r^2\right>$ ($\simeq 0.2\:\mathrm{fm}^2$), the 
numerical value of the matrix element ${\cal M}_\pi^{\mathrm{NR}}$ 
is very close to unity. If the constituent 
strange and charm quarks are treated as Dirac particles, then the matrix 
element~(\ref{nrpimat}) for $\pi^0$ emission by the strange constituent 
quark is modified to
\begin{eqnarray}
{\cal M}_\pi^{\mathrm{Rel}} &=& \frac{1}{\pi} \int_0^\infty dr'\,r'u(r')
\int_0^\infty dr\,r\,u(r) \int_0^\infty dP\,P^2 \int_{-1}^1 dz\: 
F_s^\pi(P,q_\pi,z) \nonumber \\
&&j_0\left(r'\sqrt{P^2+{q_\pi^2\over 16}+{Pq_\pi z\over 2}}\:\right)
j_0\left(r\sqrt{P^2+{q_\pi^2\over 16}-{Pq_\pi z\over 2}}\:\right),
\label{relpimat}
\end{eqnarray}
where the function $F_s^\pi$, which describes the strange quark 
contribution to pion emission by the $D_s^*$, which is a $1^3S_1$ state, 
to the $D_s$ ground state from the Lagrangian~(\ref{e4}) may be expressed 
as~\cite{pidec}
\begin{equation}
F_s^\pi(P,q_\pi,z) = \sqrt{\frac{(E'+m_s)(E+m_s)}{4EE'}}
\left(1 - \frac{P^2 - q_\pi^2/4}{3(E'+m_s)(E+m_s)}\right).
\label{spinfact}
\end{equation}
Here the energy factors of the strange quark are defined as 
$E=\sqrt{m_s^2+P^2+q_\pi^2/4-P q_\pi z}$ and 
$E'=\sqrt{m_s^2+P^2+q_\pi^2/4+P q_\pi z}$ respectively. In the above 
expressions, $\vec P$ is defined as $\vec P = (\vec p\,' + 
\vec p\,)/2$ in terms of the relative momenta in the center-of-momentum 
system. In realistic models for the heavy-light mesons~\cite{pidec,M1} and 
the baryons~\cite{Gloz,Gloz2} typical values of the constituent masses for 
the strange and charm quarks fall in the range $m_s$ = 400 MeV - 600 MeV 
and $m_c$ = 1.3 - 1.6 GeV, respectively. Because of the small constituent 
mass of the strange quark the value of the expression~(\ref{spinfact}) is 
expected to deviate significantly from the nonrelativistic 
limit~(\ref{nrpimat}). 

Because of the small value of the momentum of the $\pi^0$, one may to a 
very good approximation, as in eq.~(\ref{nrpimat}), set $q_\pi$ to zero in 
the expression~(\ref{spinfact}), which then reduces to
\begin{equation}
F_s^\pi(P) = \lim_{q\rightarrow 0}F_s^\pi(P,q_\pi,z) = 
\frac{1}{3} + \frac{2}{3}\frac{m_s}{E}.
\label{spinfact2}
\end{equation}
In this approximation the relativistic matrix element for $\pi^0$ decay 
may then be expressed as
\begin{equation}
{\cal M}_\pi^{\mathrm{Rel}} = \frac{2}{\pi} \int_0^\infty dr'\,r'u(r')
\int_0^\infty dr\,r\,u(r) \int_0^\infty dP\,P^2 \: F_s^\pi(P) \:
j_0\left(r'P\right)j_0\left(rP\right).
\label{e13}
\end{equation}
Eq.~(\ref{spinfact2}) thus indicates that in the ultrarelativistic limit, 
the matrix element for $\pi^0$ decay assumes the value $1/3$. 
For $m_s = 560$ MeV, the value of the matrix element~(\ref{e13}) 
is $\sim 0.8$, which shows that the strange constituent quark has to be 
treated relativistically. In addition, two-quark mechanisms associated 
with intermediate negative energy quarks were shown in ref.~\cite{pidec} 
to give a small contribution to the matrix element for pion decay. These 
effects will be discussed in more detail in section~\ref{3sec}.

In order to obtain the matrix element for pion emission by 
the charm quark in the $D_s^*$ meson, it is sufficient to make the 
substitution $m_s\rightarrow m_c$ in all equations of this subsection. On 
the other hand, the matrix element for the decay $\psi'\rightarrow 
J/\psi\,\eta$ takes the form
\begin{eqnarray}
{\cal M}_\eta^{\mathrm{Rel}} &=& \frac{1}{\pi} \int_0^\infty 
dr'\,r'u_{J/\psi}(r')
\int_0^\infty dr\,r\,u_{\psi'}(r) \int_0^\infty dP\,P^2 \int_{-1}^1 dz\: 
F_c^\eta(P,q_\eta,z) \nonumber \\
&&j_0\left(r'\sqrt{P^2+{q_\eta^2\over 16}+{Pq_\eta z\over 2}}\:\right)
j_0\left(r\sqrt{P^2+{q_\eta^2\over 16}-{Pq_\eta z\over 2}}\:\right),
\label{Meta}
\end{eqnarray}
where $u_{J/\psi}$ and $u_{\psi'}$ denote the reduced radial
wavefunctions for the $J/\psi$ and $\psi'$ states, respectively. The 
factor $F_c^\eta$ may be obtained by substitution of $m_s \rightarrow m_c$ 
and $q_\pi \rightarrow q_\eta$ in eq.~(\ref{spinfact}). 

\subsection{Matrix element for $\gamma$ decay}
\label{gammasec}

In the nonrelativistic approximation the matrix element for radiative M1 
decay of a $c\bar s$ meson in eq.~(\ref{rat}) is~\cite{M1}
\begin{equation}
{\cal M}_\gamma^{\mathrm{NR}} = \frac{1}{12}\left[\frac{2}{m_c} - 
\frac{1}{m_s}\right].
\label{nrgamma}
\end{equation}
Although the photon momentum $q_\gamma$ is somewhat larger 
(139 MeV) than $q_\pi$, it is still small enough so that the M1 
approximation may be considered valid. If the strange and charm 
constituent quarks are treated as Dirac particles, then the canonical 
boosts lead to relativistic modifications of the spin-flip magnetic moment 
operators~\cite{M1}. The relativistic matrix element for M1 decay may 
thus, in analogy with eqs.~(\ref{relpimat}) and~(\ref{e13}), be written 
as
\begin{equation}
{\cal M}_\gamma^{\mathrm{Rel}} = \frac{2}{\pi} \int_0^\infty dr'\,r'u(r')
\int_0^\infty dr\,r\,u(r) \int_0^\infty dP\,P^2 \: F_s^\gamma(P) \:
j_0\left(r'P\right)j_0\left(rP\right),
\label{relgmat}
\end{equation}
where the factor $F_s^\gamma$ is defined as
\begin{equation}
F_s^\gamma(P) = \frac{1}{12}\left[\frac{2}{m_c}f_c^\gamma(P) - 
\frac{1}{m_s}f_s^\gamma(P)\right].
\label{gfact}
\end{equation}
The functions $f_i^\gamma(P)$ represent correction factors to the static 
spin-flip magnetic moment operators, and may be expressed as
\begin{equation}
f_i^\gamma(P) = \frac{m_i}{3E_i}\left[2 + 
\frac{m_i}{E_i}\right] \label{fcg}
\end{equation}
for the charm and strange constituent quarks. Here the energy factors are 
defined as $E_i=\sqrt{P^2+m_i^2}$. From eq.~(\ref{nrgamma}) it is evident 
that there is destructive interference between the contributions from the 
heavy and light quark currents to the M1 decay rate of a heavy-light 
meson. Insertion of standard values of $m_c \sim 1500$ MeV and $m_s \sim 
500$ MeV for the constituent quark masses reveals that the light 
constituent quark contribution is somewhat larger than that of the heavy 
quark, which leads to an overall negative value for the $\gamma$ decay 
matrix element, cf. Table~\ref{matrtab}. However, if the relativistic 
form, eq.~(\ref{relgmat}), is employed then the charm and strange quark 
contributions very nearly cancel each other. This occurs because of the 
larger relativistic suppression of the strange quark contribution. The end 
result is, that in the relativistic impulse approximation the width for M1 
decay of a charged heavy-light meson is very small~\cite{M1}.

This result is, however, unrealistic, since two-quark mechanisms 
associated with intermediate negative energy quarks will contribute 
significantly to the matrix elements for M1 decay. These mechanisms, which 
are illustrated by the Feynman diagrams in Fig.~\ref{feyn} appear because 
of the explicit elimination of negative energy components in the 
Blankenbecler-Sugar reduction of the Bethe-Salpeter equation. The 
resulting two-quark magnetic moment operators have been 
calculated for equal constituent masses in ref.~\cite{Blunden}. It has 
been shown in ref.~\cite{oldNyf} that the two-quark contribution 
associated with the scalar confining interaction is required for obtaining 
agreement with experiment for the M1 decays in the charmonium ($c\bar c$) 
system. Without that contribution, the width for $\psi'\rightarrow 
J/\psi\:\gamma$ would be overpredicted by a factor 
$\sim 3$. 

However, the situation is much more complicated for the heavy-light 
mesons because of the uncertain structure of $Q\bar q$ interaction. 
Relativistic effects are also likely to be substantial for 
the two-quark operators because of the low masses of the light 
constituent quarks. It will be shown in the next two sections that once 
the two-quark effects associated with a $Q\bar q$ interaction formed of 
scalar confining and OGE components are taken into account, 
then the matrix elements for M1 decay of the $D^{\pm*}$ and $D_s^{\pm*}$ 
may be restored to the same order of magnitude as the non-relativistic 
estimate in eq.~(\ref{nrgamma}). The form of these two-quark operators 
will be outlined in section~\ref{3sec}, and the numerical results are 
given in section~\ref{res-sec}.


\section{Two-quark contributions to $D_s^*$ decay}
\label{3sec}

\subsection{Two-quark contributions to $\gamma$ decay}
\label{3secb}

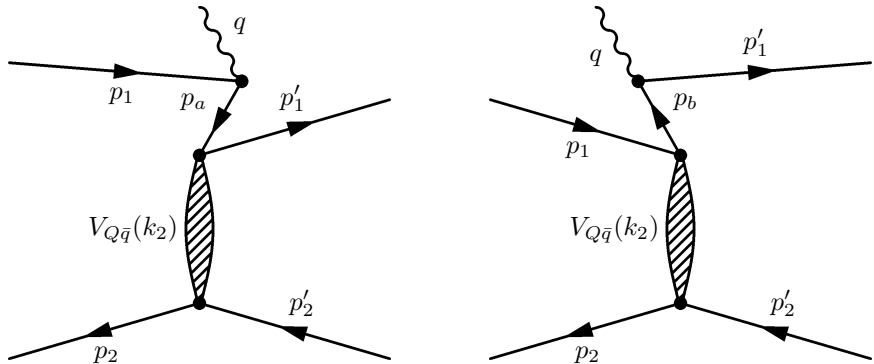
\begin{figure}[h!]
\begin{center}
\begin{tabular}{c c}
\begin{fmffile}{ex1}
\begin{fmfgraph*}(160,140) \fmfpen{thin}
\fmfcmd{%
 vardef port (expr t, p) =     
  (direction t of p rotated 90)
   / abs (direction t of p)
 enddef;}
\fmfcmd{%
 vardef portpath (expr a, b, p) =
  save l; numeric l; l = length p;
  for t=0 step 0.1 until l+0.05:
   if t>0: .. fi point t of p
    shifted ((a+b*sind(180t/l))*port(t,p))
  endfor
  if cycle p: .. cycle fi
 enddef;}
\fmfcmd{%
 style_def brown_muck expr p =
  shadedraw(portpath(thick/2,2thick,p)    
   ..reverse(portpath(-thick/2,-2thick,p))
   ..cycle)
 enddef;}
\fmfleft{i2,i1}
\fmfright{o2,o1}
\fmftop{o3}
\fmf{fermion,label=$p_1$}{i1,v3} 
\fmf{fermion,label=$p_a$}{v3,v1}  
\fmf{fermion,label=$p_1'$}{v1,o1}
\fmf{fermion,label=$p_2'$}{o2,v2}
\fmf{fermion,label=$p_2$}{v2,i2}
\fmf{photon,label=$q$}{v3,o3}
\fmf{brown_muck,lab.s=right,lab.d=4thick,lab=$V_{Q\bar q}(k_2)$}{v1,v2}
\fmfdot{v1,v2,v3}
\fmfforce{(.05w,.85h)}{i1}
\fmfforce{(.95w,.75h)}{o1}
\fmfforce{(.05w,.05h)}{i2}
\fmfforce{(.95w,.05h)}{o2}
\fmfforce{(.5w,.60h)}{v1}
\fmfforce{(.5w,.20h)}{v2}
\fmfforce{(.6w,.80h)}{v3}
\fmfforce{(.5w,.h)}{o3}
\end{fmfgraph*}
\end{fmffile}
&
\begin{fmffile}{ex2}
\begin{fmfgraph*}(160,140) \fmfpen{thin}
\fmfcmd{%
 vardef port (expr t, p) =   
  (direction t of p rotated 90)
   / abs (direction t of p)
 enddef;}
\fmfcmd{%
 vardef portpath (expr a, b, p) =
  save l; numeric l; l = length p;
  for t=0 step 0.1 until l+0.05:
   if t>0: .. fi point t of p
    shifted ((a+b*sind(180t/l))*port(t,p))
  endfor
  if cycle p: .. cycle fi
 enddef;}
\fmfcmd{%
 style_def brown_muck expr p =
  shadedraw(portpath(thick/2,2thick,p)
   ..reverse(portpath(-thick/2,-2thick,p))
   ..cycle)
 enddef;}
\fmfleft{i2,i1}
\fmfright{o2,o1}
\fmftop{o3}
\fmf{fermion,label=$p_1$}{i1,v1}
\fmf{fermion,label=$p_b$,label.side=right}{v1,v3}
\fmf{fermion,label=$p_1'$}{v3,o1}
\fmf{fermion,label=$p_2'$}{o2,v2}
\fmf{fermion,label=$p_2$}{v2,i2}
\fmf{photon,label=$q$}{v3,o3}
\fmf{brown_muck,lab.s=right,lab.d=4thick,lab=$V_{Q\bar q}(k_2)$}{v1,v2}
\fmfdot{v1,v2,v3}
\fmfforce{(.05w,.75h)}{i1}
\fmfforce{(.95w,.85h)}{o1}
\fmfforce{(.05w,.05h)}{i2}
\fmfforce{(.95w,.05h)}{o2}
\fmfforce{(.5w,.60h)}{v1}
\fmfforce{(.5w,.20h)}{v2}
\fmfforce{(.4w,.80h)}{v3}
\fmfforce{(.3w,.h)}{o3}
\end{fmfgraph*}
\end{fmffile} \\
\end{tabular}
\caption{Negative energy Born diagrams for photon emission by a $c\bar s$ 
meson. The diagrams shown correspond to both time orderings of the $\gamma$ 
emission from the charm quark. Two similar diagrams describe $\gamma$ 
emission by the strange antiquark. The quark momenta are defined according 
to $p_a = p_1+q$ and $p_b = p_1'-q$, while $k_2$ is the momentum transfered 
to the strange antiquark. Note that only the negative energy component of 
the intermediate quark propagator is to be retained.}
\label{feyn}
\end{center}
\end{figure}

In the reduction of the Bethe-Salpeter equation for the $s\bar c$
system to a Blankenbecler-Sugar equation, the negative energy
components are eliminated in the impulse approximation. These
do however contribute to the current matrix elements of the $s\bar c$
system as transition matrix elements, as illustrated in 
Fig.~\ref{feyn}. In the Blankenbecler-Sugar equation framework
these transition matrix elements have to be included as
explicit two-quark current operators \cite{Coester}. 
These two-quark matrix elements have been shown 
to give a large correction to the spin-flip matrix element of the single 
quark current~\cite{oldNyf}. 

The qualitative effect of this two-quark current may be understood in 
terms of a shift of the constituent quark mass by the scalar confining 
interaction $m\rightarrow m+V_c(r)$. Since the constituent mass appears in 
the denominator of the magnetic moment operator, this mass shift will lead 
to a reduction of the corresponding single quark contribution to the M1 
decay width. However, because of the cancellation of the relativistic 
single quark current matrix element~(\ref{relgmat}) for the $c\bar s$ 
system outlined in section~\ref{gammasec}, the two quark currents 
give the main contribution to the transition rate for $D_s^*\rightarrow 
D_s\gamma$~\cite{M1}. For the scalar confining interaction, that current 
may be expressed in the form
\begin{eqnarray}
\vec\jmath_c(\vec q,\vec k_1,\vec k_2) = -e\,\left(
\frac{Q_1^*\vec P_1}{m_1^2} + \frac{Q_2^*\vec 
P_2}{m_2^2}\right.\!\!\!\!\!\!\!\!\!\!\!
&&\left.+\:\:\frac{i}{2}(\vec\sigma_1+\vec\sigma_2)\times\vec q\left[
\frac{Q_1^*}{2m_1^2} + \frac{Q_2^*}{2m_2^2}\right]\right. \nonumber \\
&&\left.+\:\:\frac{i}{2}(\vec\sigma_1-\vec\sigma_2)\times\vec q\left[   
\frac{Q_1^*}{2m_1^2} - \frac{Q_2^*}{2m_2^2}\right]\right), \label{2qcurr}
\end{eqnarray}
where $\vec q$ is the momentum of the emitted photon, which corresponds to 
$\vec p\,' - \vec p$ in case of the single quark amplitude. 
In eq.~(\ref{2qcurr}), the variables $Q_1^*$ and $Q_2^*$ are defined as
$Q_1^* = V_c(\vec k_2)Q_1$ and $Q_2^* = V_c(\vec k_1)Q_2$, where $V_c(\vec 
k_n)$ is the (formal) Fourier transform of the scalar confining 
interaction. The momentum 
variables $\vec P_1$ and $\vec P_2$ correspond to the expressions $(\vec 
p_1 + \vec p\,'_1)/2$ and $(\vec p_2 + \vec p\,'_2)/2$ respectively, while 
$\vec k_n$ denotes the momentum transfered to quark $n$ according to 
Fig.~\ref{feyn}. The analogous expression for the two-quark current induced 
by the vector coupled OGE interaction is of the form
\begin{eqnarray}
\vec\jmath_g(\vec q,\vec k_1,\vec k_2) \!\!&=&\!\! -e\,\left(Q_1^*\left[
\frac{i\vec\sigma_1\times\vec k_2}{2m_1^2} + 
\frac{2\vec P_2 + i\vec\sigma_2\times\vec k_2}{2m_1m_2}\right] 
+Q_2^*\left[
\frac{i\vec\sigma_2\times\vec k_1}{2m_2^2} + 
\frac{2\vec P_1 + i\vec\sigma_1\times\vec k_1}{2m_1m_2}\right]\right),
\label{Ogecurr}
\end{eqnarray}
with $Q_1^* = V_g(\vec k_2)Q_1$ and $Q_2^* = V_g(\vec k_1)Q_2$. Here 
$V_g(\vec k)$ denotes the momentum-space form of the OGE interaction. The 
corresponding magnetic moment operators may then be computed from 
eqs.~(\ref{2qcurr}) and~(\ref{Ogecurr}) according to
\begin{equation}
\vec\mu\equiv -\frac{i}{2}\lim_{q\rightarrow 0}\left[\vec\nabla_q\times 
\vec\jmath\:(\vec q,\vec k_1,\vec k_2)\right],
\end{equation}
where the momentum transfer variables are given by $\vec k_1 = \vec q/2 - 
\vec k$ and $\vec k_2 = \vec q/2 + \vec k$. 
Upon Fourier transformation, the resulting magnetic moment operators may, 
for transitions between $S$-wave states, be 
written in the form
\begin{equation}
\vec\mu_c = -\frac{eV_c(r)}{4}\left\{\left[\frac{Q_1}{m_1^2} - 
\frac{Q_2}{m_2^2}\right]\:(\vec\sigma_1 - \vec\sigma_2)
+\left[\frac{Q_1}{m_1^2} +
\frac{Q_2}{m_2^2}\right]\:(\vec\sigma_1 + \vec\sigma_2)\right\}
\label{muc}
\end{equation}
for the scalar confining interaction, and
\begin{equation}
\vec\mu_g = -\frac{eV_g(r)}{8}\left\{\left[\frac{Q_1}{m_1^2} - 
\frac{Q_2}{m_2^2}
- \frac{Q_1-Q_2}{m_1m_2}\right]\:(\vec\sigma_1 - \vec\sigma_2)
+\left[\frac{Q_1}{m_1^2} + \frac{Q_2}{m_2^2}
+ \frac{Q_1+Q_2}{m_1m_2}\right]\:(\vec\sigma_1 + \vec\sigma_2)\right\}
\label{mug}
\end{equation}
for the one-gluon exchange interaction. In the limit $m_1 = m_2$, the 
above magnetic moments reduce to those given in ref.~\cite{Blunden}. Note 
that for the case of equal constituent quark masses, the one-gluon 
exchange magnetic moment operator has no spin-flip term and 
consequently does not contribute to M1 decay if $m_1 = m_2$. The terms in 
eqs.~(\ref{muc}) and~(\ref{mug}) that are symmetric in the quark spins 
affect only the magnetic moments of the heavy-light mesons and give no 
contribution to the M1 decay widths.

The spin-flip 
matrix elements for radiative M1 decay of a $c\bar s$ meson 
that should be added to the relativistic single quark matrix 
element~(\ref{relgmat}) are thus
\begin{equation}
{\cal M}_\gamma^{\mathrm{Conf}} = -\int_0^\infty 
dr\,u^2(r)\,\frac{V_c(r)}{12}\left[\frac{2}{m_c^2} - 
\frac{1}{m_s^2}\right],
\label{NRConf}
\end{equation}
in case of the scalar confining interaction, and
\begin{equation}
{\cal M}_\gamma^{\mathrm{Oge}} = -\int_0^\infty
dr\,u^2(r)\,\frac{V_g(r)}{24}\left[\frac{2}{m_c^2} -
\frac{1}{m_s^2} - \frac{1}{m_c m_s}\right]
\label{NROge}
\end{equation}
for the one-gluon exchange interaction. In the above expressions, the 
appropriate values of the quark charge operators have been entered. For 
the scalar confining interaction, the form $V_c(r) = cr - b$ will be used, 
where the parameters $c$ and $b$ are taken to be those obtained in the 
potential model of ref.~\cite{M1}. The one-gluon exchange potential 
$V_g(r)$ is taken to be of the form
\begin{equation}
V_g(r) = -\frac{4}{3}\frac{2}{\pi}\int_0^\infty dk\:\alpha_s(\vec k^2) 
\,j_0(kr),
\end{equation}
which is known as the Richardson potential~\cite{Davies} and features the 
running QCD coupling $\alpha_s$ which is parameterized in terms of the QCD 
scale 
$\Lambda_{\mathrm{QCD}}$ and a dynamical gluon mass $m_g$, which in 
ref.~\cite{M1} were obtained as 280 MeV and 240 MeV, respectively:
\begin{equation}
\alpha_s(k^2)=\frac{12\pi}{27}\frac{1}
{\ln [(k^2+4m_g^2)/\Lambda_\mathrm{QCD}^2]}.\label{zzz}
\end{equation}
Note that if $\alpha_s$ is taken to be constant, then the above form 
reduces to the standard Coulombic one-gluon exchange potential of 
perturbative QCD.

\subsection{Two-quark contributions to $\pi^0$ decay}
\label{3seca}

In ref.~\cite{pidec} it was shown that a two-quark mechanism analogous to 
that for M1 decay shown in Fig.~\ref{feyn}, which is 
due to coupling of the emitted pion to an intermediate negative energy 
quark, which interacts with the heavy spectator quark by the $Q\bar q$ 
interaction may contribute to the amplitude for pion decay. If the decay 
amplitude for emitted pions is written in the form
\begin{equation}
T_\pi = i\vec q_\pi \cdot \vec A,
\end{equation}
then the axial current $\vec A$ may be decomposed into 
single- and two-quark contributions according to $\vec A = \vec A_s + \vec 
A_{\mathrm {ex}}$. The two-quark contributions, which have been calculated 
for equal constituent masses in ref.~\cite{Tsushima}, are in this case 
however proportional to $m_q^{-3}$ in the nonrelativistic limit, which 
means that they are of much less importance than the two-quark 
contributions to the magnetic moment operator. Furthermore, it has been 
shown in ref.~\cite{pidec} that the axial exchange current contribution 
associated with the scalar confining interaction is much reduced if 
relativistic effects are considered, because of the low mass of the 
strange constituent quark.

The end result is, that two-quark effects due to the scalar confining 
and one-gluon exchange interactions will only modify the 
relativistic matrix element~(\ref{e13}) at the $\sim 5-10$\% level. As the 
uncertainties in the value of the mixing angle $\theta_m$ and the 
empirically determined ratio of $\gamma$ and $\pi^0$ decay are much 
greater, there is no need to include the two-quark contributions to the 
amplitude for $\pi^0$ decay at this time.

\newpage

\section{Numerical Results for $D_s^*\rightarrow D_s\gamma$ and 
$D_s^*\rightarrow D_s\pi^0$}
\label{res-sec}

In 
order to obtain a description of the decay $D_s^*\rightarrow D_s\gamma$ 
that is as realistic as possible, the model for M1 decay presented here 
should be checked against the empirical results on the corresponding 
decays of the non-strange $D$ mesons. As the decay width for 
$D^{\pm *}\rightarrow D^\pm \gamma$ is known directly and that for 
$D^{0*}\rightarrow D^0\gamma$ can be estimated from the empirical 
branching ratios and model calculations of the pionic decays of the $D$ 
mesons, it is possible to calibrate the mass of the light constituent 
quark so that an optimal description of the M1 decay of the $D^\pm$ is 
obtained. Together with the 
potential model calculation in ref.~\cite{M1}, this information can 
then be used to obtain a realistic estimate of the strange quark 
constituent mass. In order to calculate the M1 decay rate $D^{\pm 
*}\rightarrow D^\pm \gamma$, it is sufficient to make the substitution 
$m_s\rightarrow m_q$ in all matrix elements, but for $D^{0*}\rightarrow 
D^0\gamma$, one should substitute $m_s^{-1}\rightarrow -2m_q^{-1}$ in 
eq.~(\ref{gfact}), and $m_s^{-2}\rightarrow -2m_q^{-2}$, $1/m_c 
m_s\rightarrow 4/m_cm_q$ in eqs.~(\ref{NRConf}) and~(\ref{NROge}). The 
obtained results for the M1 widths of the non-strange $D$ mesons are given 
for different light constituent quark masses in Tables~\ref{Dpmtab} 
and~\ref{D0tab}.
\vspace{.5cm}
\begin{table}[h!]
\centering{
\begin{tabular}{c|c|c|c}
$m_q$ & NRIA 	& RIA 		     & RIA + Conf + Oge 	\\ 
\hline\hline
&&& \\ 
450 MeV & 0.58 keV	& $9.4\cdot 10^{-3}$ keV & 1.09 keV	\\
&&& \\
420 MeV & 0.79 keV	& $1.5\cdot 10^{-2}$ keV & 1.43	keV	\\
&&& \\
390 MeV & 1.07 keV	& $2.2\cdot 10^{-2}$ keV & 1.90	keV	\\
\end{tabular}
\caption{Numerical results for the M1 transition $D^{\pm *}\rightarrow 
D^\pm \gamma$, for a charm quark mass of 1580 MeV~\cite{M1}. The column 
NRIA (non-relativistic impulse approximation) corresponds to 
eq.~(\ref{nrgamma}), and the relativistic impulse 
approximation RIA to eq.~(\ref{relgmat}). In the rightmost column, the 
two-quark contributions from eqs.~(\ref{NRConf}) and~(\ref{NROge}) have 
been added. In the contribution from the confining interaction, the 
parameters $c$ = 1120 MeV/fm and $b$ = 320 MeV have been 
employed~\cite{M1}.}\label{Dpmtab}}
\end{table}

\begin{table}[h!]
\centering{
\begin{tabular}{c|c|c|c}
$m_q$ & NRIA 		& RIA		& RIA + Conf + Oge 	\\ 
\hline\hline
&&& \\ 
450 MeV & 21.1 keV	& 8.86 keV 	& 8.95 keV	\\
&&& \\
420 MeV & 23.5 keV	& 9.18 keV 	& 9.89 keV	\\
&&& \\
390 MeV & 26.4 keV	& 9.52 keV 	& 11.1 keV	\\
\end{tabular}
\caption{Numerical results for the M1 transition $D^{0*}\rightarrow 
D^0 \gamma$, for a charm quark mass of 1580 MeV~\cite{M1}. The columns 
NRIA (non-relativistic impulse approximation) and 
RIA are as for Table~\ref{Dpmtab}, although with the appropriate 
modifications due to the different quark charge operators. The same 
modifications apply to the rightmost column, where the 
two-quark contributions from eqs.~(\ref{muc}) and~(\ref{mug}) have 
been added. In the contribution from the confining interaction, the 
parameters $c$ = 1120 MeV/fm and $b$ = 320 MeV have been 
employed~\cite{M1}.}\label{D0tab}}
\end{table}

\newpage

It is instructive to compare the results presented in Table~\ref{Dpmtab} 
with the current experimental data for the $D^{\pm *}$. Until recently, 
only the relative branching ratios for $\pi$ and $\gamma$ decay were 
known~\cite{PDG}. However, a first empirical measurement of the total 
width of the $D^{\pm *}$ has recently been published by the CLEO 
collaboration~\cite{Cleo}. The reported result is $\Gamma(D^{\pm *}) = 96 
\pm 4 \pm 22$ keV, where the latter error represents the systematic 
uncertainties. Taking into account the reported~\cite{PDG} branching ratio 
of $1.6\pm 0.4$\% for radiative decay, one obtains $1.5\pm 0.6$ keV for 
the transition $D^{\pm *}\rightarrow D^\pm \gamma$. Here most of the 
uncertainty stems from the systematic errors of the experimental result. 
However, it is evident that the results of Table~\ref{Dpmtab} reproduce 
this result well for a range of values of the light constituent quark mass 
$m_q$. The value $m_q = 450$ MeV corresponds to the potential model of 
ref.~\cite{M1}, while the value $m_q = 420$ MeV has been suggested by 
refs.~\cite{ch1,ch2}. From Table~\ref{Dpmtab} it is thus seen that a light 
constituent quark mass of 420 MeV produces a width for radiative decay 
which is close to 1.5 keV. That value of the width is also favored by the 
analysis of ref.~\cite{GoityM1}.

\begin{table}[h!]
\centering{
\begin{tabular}{c|c|c|c}
$m_q$ & NRIA 	& RIA 		     & RIA + Conf + Oge 	\\ 
\hline\hline
&&& \\ 
560 MeV & 0.18 keV	& $2.6\cdot 10^{-4}$ keV & 0.38 keV	\\
&&& \\
530 MeV & 0.26 keV	& $3.9\cdot 10^{-5}$ keV & 0.49	keV	\\
&&& \\
500 MeV & 0.36 keV	& $8.6\cdot 10^{-4}$ keV & 0.64	keV	\\
\end{tabular}
\caption{Numerical results for the M1 transition $D_s^{\pm *}\rightarrow 
D_s^\pm \gamma$, for a charm quark mass of 1580 MeV~\cite{M1}. The column 
NRIA (non-relativistic impulse approximation) corresponds to 
eq.~(\ref{nrgamma}), and the relativistic impulse 
approximation RIA to eq.~(\ref{relgmat}). In the rightmost column, the 
two-quark contributions from eqs.~(\ref{NRConf}) and~(\ref{NROge}) have 
been added. In the contribution from the confining interaction, the 
parameters $c$ = 1120 MeV/fm and $b$ = 260 MeV have been 
employed~\cite{M1}. Note that the RIA contribution becomes exactly zero 
for a strange quark mass of $\sim 540$ MeV.}\label{Dstab}}
\end{table}
\vspace{.4cm}

Even though the total width of the neutral $D^{0*}$ meson has not yet been 
determined~\cite{PDG}, considerable information about the expected width 
for $D^{0*}\rightarrow D^0 \gamma$ may be extracted from the reported 
branching fraction of $38.1 \pm 2.9$\%~\cite{PDG}, since the corresponding 
width for pion decay can be constrained by means of the empirically 
determined width of the $D^{\pm *}$ and model calculations of the pionic 
decays of $D$ mesons~\cite{pidec,Goity,Eichten}. This is seen directly if 
one first notes that the branching fraction of $D^{\pm *}\rightarrow D^0 
\pi^\pm$ is reported as $67.7\pm 0.5$\%~\cite{PDG}, which implies a width 
for this decay mode of $\sim 65 \pm 14$ keV. From the model 
calculation of ref.~\cite{pidec} one finds that this corresponds to a 
width of $\sim 40\pm 10$ keV for $D^{0*}\rightarrow D^0 \pi^0$. As the 
relative branching fractions for 
$\pi^0$ and $\gamma$ decay of the $D^{0*}$ are well known~\cite{PDG}, the 
best estimate for the width of $D^{0*}\rightarrow D^0 \gamma$ is $\sim 25$ 
keV. There remains, however, still a considerable uncertainty of $\sim \pm 
10$ keV from the systematic errors in the empirical measurement of 
$\Gamma(D^{\pm *})$. 

A width for $D^{0*}\rightarrow D^0 \gamma$ of around 
20-30 keV is also preferred by ref.~\cite{GoityM1}. The result of 
the model calculation in Table~\ref{D0tab} is smaller by about a 
factor~$\sim 2$ as the OGE contribution is weaker in this case as compared 
to the contribution from the scalar confining interaction. Moreover, as a 
relativistic treatment of the two-quark currents will effectively
lead to replacement of the quark masses by energy-dependent factors, it   
may be argued that relativistic effects will generally lead to a weakening
of the two-quark contributions to the matrix element for M1 decay. It is
thus entirely possible that a fully relativistic treatment will render the
two-quark contributions too weak to account for the experimental data on
M1 decay of the $D^\pm$ meson as well.

This conclusion is in line with that reached in ref.~\cite{GoityM1}, where 
it was found that the introduction of a large anomalous magnetic moment is 
required in order to fit the observed M1 decay widths of the heavy-light 
mesons. Another interesting possibility is that the instanton induced 
interaction for $Q\bar q$ systems given in refs.~\cite{ch1,ch2} may also 
contribute a significant two-quark current. The interaction of 
refs.~\cite{ch1,ch2}, which was found to be short-ranged, negative and 
attractive, has scalar coupling to the light constituent quark. One may 
thus conclude that this interaction will add up constructively with the OGE 
contribution from eq.~(\ref{NROge}). Thus the instanton 
induced interaction will have an overall favorable effect on the widths 
for M1 decay, particularly for the $D^{0*}$, which may be inferred from 
the matrix elements given in Table~\ref{matrtab}.

As it appears to be possible to obtain a realistic description of the 
radiative M1 decay of the $D^\pm$ meson using the model presented in this 
paper, then a prediction based on the potential model~\cite{M1} for the M1 
decay $D_s^{\pm *}\rightarrow D_s^\pm \gamma$ and the associated matrix 
element ${\cal M}_\gamma$ can be made. That information is needed as 
input for eq.~(\ref{rat}) in order to obtain an estimate for $f_{\eta ss}$. 
These results are given in Tables~\ref{Dstab} and~\ref{matrtab}.

\begin{table}[h!]
\centering{
\begin{tabular}{l||r|r|r}
 & $D^{\pm *}\rightarrow D^\pm \gamma$ & $D^{0*}\rightarrow D^0 \gamma$
 & $D_s^{\pm *}\rightarrow D_s^\pm \gamma$ \\ \hline\hline &&& \\
${\cal M}_\gamma^{\mathrm{NR}}$ & $-1.83\cdot 10^{-2}$ & $9.91\cdot 10^{-2}$
 & $-8.55\cdot 10^{-3}$ \\ &&& \\
${\cal M}_\gamma^{\mathrm{Rel}}$& $-2.55\cdot 10^{-3}$ & $6.20\cdot 10^{-2}$
 & $ 3.24\cdot 10^{-4}$ \\ &&& \\
${\cal M}_\gamma^{\mathrm{Conf}}$&$ 1.23\cdot 10^{-2}$ & $-3.07\cdot 10^{-2}$
 & $ 7.25\cdot 10^{-3}$ \\ &&& \\
${\cal M}_\gamma^{\mathrm{Oge}}$& $-3.45\cdot 10^{-2}$ & $3.31\cdot 10^{-2}$
 & $ -1.98\cdot 10^{-2}$ \\ &&& \\ \hline &&& \\
${\cal M}_\gamma^{\mathrm{Tot}}$& $-2.48\cdot 10^{-2}$ & $6.44\cdot 10^{-2}$ 
 & $ -1.22\cdot 10^{-2}$ \\
\end{tabular}
\caption{Matrix elements for the M1 decays considered in 
Tables~\ref{Dpmtab},~\ref{D0tab} and~\ref{Dstab}, in units of [fm]. The 
matrix elements correspond to a charm quark mass of 1580 MeV, a light 
constituent quark mass of 420 MeV and a strange constituent quark mass of 
560 MeV. The matrix element ${\cal M}_\gamma^{\mathrm{Tot}}$ shows the sum 
of the 'Rel' + 'Conf' + 'Oge' contributions, which in case of the decay 
$D_s^{\pm *}\rightarrow D_s^\pm \gamma$ is used for the determination of 
the coupling constant $f_{\eta ss}$.}\label{matrtab}}
\end{table}

In order to estimate the coupling constant $f_{\eta ss}$, the matrix 
element~(\ref{e13}) for $\pi^0$ decay of the $D_s^*$ meson also needs to be 
evaluated. The value of that matrix element for a strange constituent quark 
mass of 560 MeV is 0.794, which together with the total 
matrix element for $\gamma$ decay of the $D_s^*$ may then be used with 
eq.~(\ref{rat}) to obtain a value for $f_{\eta ss}$. If the charm quark 
also couples significantly to the $\eta$ meson, then the matrix element for 
pion emission by the charm quark in eq.~(\ref{req}) needs to be evaluated 
as well. This can be done by the substitution $m_s\rightarrow m_c$ in 
eq.~(\ref{e13}). For a charm quark mass of 1580 MeV, the result ${\cal 
M}_\pi^c$ = 0.949 is obtained. Thus, in the evaluation of eqs.~(\ref{Req}) 
and~(\ref{req}), the following values for the matrix elements for $\gamma$ 
and $\pi^0$ decay will be used:
\vspace{0.3cm}
\begin{center}
\fbox{${\cal M}_\gamma = -1.22\cdot 10^{-2}\:\mathrm{fm}$}
\end{center}
\vspace{0.3cm}   
for the $\gamma$ decay matrix element, and
\vspace{0.3cm}
\begin{center}
\fbox{${\cal M}_\pi^s = 0.794\quad\quad {\cal M}_\pi^c = 0.949$}
\end{center}
\vspace{0.3cm}   
for the $\pi^0$ decay matrix elements. Note that in all the calculations of 
this section, the 
masses of the initial and final quarkonium states as well as the momenta 
of the emitted photons and $\pi^0$ mesons have been taken to 
equal those given in ref.~\cite{PDG}.

\subsection{Numerical Results for $\psi'\rightarrow J/\psi\:\eta$}

The evaluation of eq.~(\ref{etawidth}) involves the computation of the 
matrix element ${\cal M}_\eta$ given by eq.~(\ref{Meta}). The $\psi'$ 
and $J/\psi$ wavefunctions are here taken to be those computed in the 
potential model of ref.~\cite{Resc}. Because of the somewhat larger value 
of $q_\eta$~(200 MeV)~\cite{PDG}, and the orthogonality of the $\psi'$ and 
$J/\psi$ radial wavefunctions, it is undesirable to make the approximation 
$q_\eta = 0$ in this case. The resulting relativistic matrix element for 
the decay $\psi'\rightarrow J/\psi\:\eta$ is obtained as ${\cal M}_\eta = 
-3.53\cdot 10^{-2}$, and gives the desired width of $\sim 7.5$ keV if the 
coupling constant $f_{\eta cc}$ is taken to be $\sim 0.8$. The sign of 
$f_{\eta cc}$ is of course not determined by this computation. 

As there exists both theoretical and phenomenological 
arguments~\cite{Feldmann} against a significant $c\bar c$ admixture in the 
$\eta$ meson, then this value for $f_{\eta cc}$ may appear 
uncomfortably large. Indeed, it will be shown in the next section that a 
value of $f_{\eta cc}$ around $\sim 0.8$ is, in view of the 
$D_s^*\rightarrow D_s\pi^0$ decay, about one half of the so obtained value 
for $f_{\eta ss}$. However, as neither the axial charge component nor the 
associated two-quark contributions to the amplitude for $\eta$ decay of the 
$\psi'$ have been considered, the value of 0.8 for $f_{\eta cc}$ should be 
considered as an upper limit only. Nevertheless, it will be shown in the 
next section that the inclusion of a pion-charm quark coupling will serve 
to increase the obtained value for $f_{\eta ss}$, thus bringing it closer 
to the phenomenologically preferred values~\cite{Tiator,Pena}.

\section{Discussion}
\label{4sec}

Assuming that in the $\pi^0$ decay of the $D_s^*$, the pion couples mostly 
to the strange constituent quark, eq.~(\ref{rat}) may be used directly 
together with the matrix elements for $\pi^0$ and $\gamma$ decay given in 
the previous section to estimate the magnitude of $f_{\eta ss}$. Insertion 
of the matrix elements and taking $\theta_m = 0.012$ yields $R = 0.3085$, 
giving $|f_{\eta ss}| = 0.70$. Taking into account the uncertainties in 
the mixing angle and the empirical decay widths for $\pi^0$ and $\gamma$ 
decay, the best estimate is
\vspace{0.1cm}
\begin{center}
\fbox{$ f_{\eta ss}\:\:=\:\: -\,0.7\:^{+0.5}_{-0.3}$}
\end{center}
\vspace{0.1cm}
In the above result, the negative sign is suggested by the relations 
in eq.~(\ref{Treimanq}). The static quark model then implies, through 
eq.~(\ref{qmod}), that the magnitude of the corresponding pseudovector 
$\eta$-nucleon coupling constant $f_{\eta NN}$ should be one half of this 
value. Thus one obtains the following value for the $\eta$-nucleon 
coupling:
\vspace{0.1cm}
\begin{center}
\fbox{$ f_{\eta NN}\:\:=\:\: 0.35\:^{+0.15}_{-0.25}$}
\end{center}
\vspace{0.1cm}
This result should be compared with the value for $f_{\eta NN}$ (or the 
equivalent pseudoscalar coupling constant $g_{\eta NN} = (2 m_N/m_\eta) 
f_{\eta NN}$, which has been determined by phenomenological model fits to 
photoproduction of $\eta$ mesons on the nucleon~\cite{Tiator}. The latter 
value for $f_{\eta NN}$ is 0.64. This value has also been found to be 
realistic in calculations of the cross section for $pp\rightarrow pp\eta$ 
near threshold~\cite{Pena}. Although the result obtained above for $f_{\eta 
NN}$ has quite large uncertainties which are mostly of empirical origin, it 
still appears to be significantly smaller. A larger value for $f_{\eta NN}$ 
could of course be obtained by decreasing the mixing angle $\theta_m$.

Additional modifications to the current estimation of $f_{\eta NN}$ may of 
course also arise from $SU(3)$ breaking effects in eq.~({\ref{Treimanq}), 
which relates $f_{\eta qq}$ to $f_{\eta ss}$. Likewise, relativistic 
effects may also modify the static quark model result~(\ref{qmod}) which 
relates $f_{\eta qq}$ to $f_{\eta NN}$. However, in view of the large 
experimental uncertainty in the determination of the branching fractions 
for $\gamma$ and $\pi^0$ decay of the $D_s^*$, there appears to be little 
motivation to include such effects at this time. Another possibility is 
that the matrix element for $\gamma$ decay as given by Table~\ref{matrtab} 
is too small. However, as the computed width for $D_s^*\rightarrow D_s 
\gamma$ is already somewhat larger than that given in ref.~\cite{GoityM1}, 
this possibility may be considered unlikely. 

Another option worth considering, in view of the fact that the 
empirical detection of a considerable branching fraction for 
$\psi'\rightarrow J/\psi\,\eta$ indicates that the $\eta$ meson also 
couples to the charm quark, is that the coupling of the $\eta$ meson to 
the charm quark would be strong enough to significantly influence the 
$D_s^*\rightarrow D_s \gamma$ decay. For nonzero values of the 
coupling $f_{\eta cc}$, the coupling constant $f_{\eta ss}$ may thus be 
determined from eq.~(\ref{req}) according to
\begin{equation}
f_{\eta ss} = \frac{\sqrt{R} - f_{\eta cc}{\cal M}_\pi^c}{{\cal M}_\pi^s},
\end{equation}
where the different possibilities are $\sqrt{R} = \pm 0.5554$ and $f_{\eta 
cc} = \pm 0.8$. The most realistic of these appears to be
$\sqrt{R} = -0.5554, f_{\eta cc} = +0.8$, giving $f_{\eta ss} =
-1.66$. In this case a nonzero and positive $f_{\eta cc}$ leads to a larger 
and negative value for $f_{\eta ss}$, allowing for better agreement with 
the values of $f_{\eta NN}$ suggested above. The value for $f_{\eta NN}$ 
that corresponds to $f_{\eta cc} = +0.8$ turns out to be 
\begin{equation}
f_{\eta NN}\:\:=\:\: 0.83\:^{+0.25}_{-0.35},
\end{equation}
which is closer to the 
phenomenological estimates given by refs.~\cite{Tiator} and~\cite{Pena}. 
Although the value +0.8 for $f_{\eta cc}$ is, in view of the discussion in 
the previous section, probably much too large, there still remains a 
distinct possibility that the charm quark contributes significantly to the 
width for $D_s^*\rightarrow D_s\pi^0$. In this case it is found that values 
around $f_{\eta ss} \simeq -1.7$ are favored by an eta-charm coupling of 
$\sim +0.8$.

It is useful to compare the results obtained from the present 
phenomenological analysis with the predictions of eq.~(\ref{Treimanq}). 
With the value $g_A^q = 0.87$ for the axial coupling constant of the 
constituent quarks, eq.~(\ref{Treimanq}) 
yields the value $f_{\pi qq}=$ 0.65 for the pion-quark coupling, which is 
close to the value $f_{\pi qq} = 3/5 f_{\pi NN} \simeq 0.57$ suggested by 
the static quark model~(cf.~(\ref{e3})). The values for $f_{\eta qq}$ and 
$f_{\eta ss}$ so obtained are 1.25 and -2.5 respectively. The values for 
$f_{\eta ss}$ obtained here from the ratio of the empirical branching 
fractions for $D_s^*\rightarrow D_s \pi^0$ and $D_s^*\rightarrow D_s 
\gamma$, are much smaller than this quark model value. In the completely 
$SU(3)$ flavor symmetric limit, in which the pion and $\eta$ are 
degenerate in mass, the value for $f_{\eta ss}$ would however be equal to 
$-2/\sqrt{3} f_{\pi qq}$, which is only -0.75. The results 
from various modern phenomenological analyses are typically about twice 
this value.

Naive application of $SU(3)$ flavor symmetry at the level of a chiral 
symmetry violating pseudoscalar coupling model for pions and
$\eta$ mesons to baryons gives the result $g_{\eta NN}=1/\sqrt{3}\,g_{\pi 
NN}$. With the (now) standard value of 12.8 for $g_{\pi NN}$ this relation
would give the value 7.4 for $g_{\eta NN}$. Such a large value would 
correspond to $f_{\eta NN}$ = 2.2. While values of that magnitude are 
ruled out by $\eta$ meson photoproduction, as well as, as shown here, by 
the $\pi^0$ decay of the $D_s^*$ meson, they were employed in early 
realistic boson exchange models for the nucleon-nucleon interaction.

\newpage

\centerline{\bf Acknowledgments}
\vspace{.6cm}
We are indebted to Dr.S. Wycech for instructive discussions. DOR thanks the 
W.K. Kellogg Radiation Laboratory of the California Institute of Technology 
for its hospitality during the completion of this work. Research supported 
in part by the Academy of Finland
by grants 43982 and 54038.

\end{document}